\documentclass[12pt]{article}
\newcommand{\be}{\begin{equation}}
\newcommand{\ee}{\end{equation}}

\newcommand{\bi}{\begin{itemize}}
\newcommand{\ei}{\end{itemize}}
\newcommand{\bea}{\begin{eqnarray}}
\newcommand{\eea}{\end{eqnarray}}
\newcommand{\ba}{\begin{array}}
\newcommand{\ea}{\end{array}}

\usepackage[left=2.50cm, right=2.50cm, top=2.50cm, bottom=2.50cm]{geometry}
\usepackage[utf8]{inputenc}
\usepackage{cancel}
\usepackage[english]{babel}
\usepackage{physics}
\usepackage{hyperref}
\usepackage{amsthm}
\usepackage{mathrsfs}
\usepackage{mathtools}
\usepackage{graphicx}
\usepackage{changepage}
\usepackage{amssymb,amsmath}
\usepackage{verbatim}
\usepackage{empheq}
\usepackage{xcolor}
\usepackage{tikz}
\usepackage{float}
\usepackage{multirow}
\usepackage{multicol}
\usepackage{amsmath}
\usepackage{centernot}
\usepackage[hang,flushmargin]{footmisc} 
\usepackage[normalem]{ulem}
\usetikzlibrary{tikzmark}
\numberwithin{equation}{section}
\hypersetup{hidelinks}

\newlength{\bibitemsep}
\newlength{\bibparskip}
\let\oldthebibliography\thebibliography
\renewcommand\thebibliography[1]{%
  \oldthebibliography{#1}%
  \setlength{\parskip}{\bibitemsep}%
  \setlength{\itemsep}{\bibparskip}%
}
\begin{document}
\par
\bigskip
\Large
\noindent
{\bf 
Edge modes in Chern-Simons theory on a strip}

\bigskip
\par
\rm
\normalsize

\hrule

\vspace{1cm}

\large
\noindent
{\bf Erica Bertolini$^{1,a}$},
{\bf Michael Doyle$^{2,b}$}, 
{\bf Nicola Maggiore$^{3,4,c}$},\\
{\bf Conor Murphy$^{2,d}$},
{\bf Carlotta Piras$^{2,e}$}\\

\par

\small

\noindent$^1$ School of Theoretical Physics, Dublin Institute for Advanced Studies, 10 Burlington Road, Dublin 4, Ireland.

\noindent$^2$ Trinity College Dublin, College Green, Dublin 2, D02K8N, Ireland.

\noindent$^3$ Dipartimento di Fisica, Universit\`a di Genova, Via Dodecaneso 33, I-16146 Genova, Italy.

\noindent$^4$ Istituto Nazionale di Fisica Nucleare - Sezione di Genova, Via Dodecaneso 33, I-16146 Genova, Italy.

\smallskip

\smallskip

\vspace{1cm}

\noindent
{\tt Abstract}\\

We investigate abelian Chern-Simons gauge theory on a strip geometry with two spatial boundaries. In the presence of boundaries, gauge invariance is broken by boundary conditions, leading to physical edge excitations. By deriving the most general local boundary conditions consistent with power counting in the sense of Symanzik, we show that the bulk equations of motion determine the boundary degrees of freedom through a broken gauge Ward identity, yielding boundary Kac-Moody current algebras with opposite central charges on the two edges. The corresponding two-dimensional boundary actions are of Tomonaga-Luttinger type and describe chiral bosons propagating in opposite directions along the two boundaries. A consistency condition, interpreted as a holographic-like bulk-boundary matching, relates the
Chern-Simons coupling constant and the boundary parameters to the physical edge velocities. Within this framework, the equality and opposite sign of the two velocities in a symmetric setup follow directly from the boundary structure rather than from model-dependent assumptions about confining potentials, and the velocities are independent of the strip width. Our analysis provides a fully field-theoretic realization of bulk-boundary correspondence in Chern-Simons theory with two boundaries, with direct applications to edge physics in quantum Hall systems and related topological/hydrodynamic settings.

\vspace{\fill}

\noindent{\tt Keywords:} \\
Chern-Simons theory, Gauge theories with boundaries, Boundary conditions, Ward identities, Kac-Moody algebra, Chiral bosons.

\vspace{1cm}

\hrule
\noindent{\tt E-mail:
$^a$ebertolini@stp.dias.ie,
$^b$doylem55@tcd.ie,
$^c$nicola.maggiore@ge.infn.it,\\
$^d$murphy33@tcd.ie,
$^e$pirasc@tcd.ie
}
\newpage

\section{Introduction }

Chern-Simons (CS) gauge theories in 3D provide a paradigmatic example of a topological quantum field theory: while the bulk dynamics is purely topological, the presence of boundaries  leads to dynamical edge degrees of freedom encoded in a broken gauge Ward identity. A prominent physical realization is provided by the quantum Hall effect: it exhibits quantized conductance that is insensitive to microscopic details, and its low-energy behavior is elegantly described by CS effective field theories
\cite{Wen:1992vi,Stone:1990iw,Fradkin:1991wy,Tong:2016kpv,Maggiore:2017vjf}. In the standard treatment, a CS theory with a single boundary reproduces the well known chiral edge mode, whose direction of propagation is fixed by the sign of the CS coupling. This framework has successfully explained numerous experimental observations and has established a direct link between bulk topological invariants and boundary phenomena
\cite{Zhang:1992eu,Amoretti:2014iza,Blasi:2019wpq}.
From a field-theoretic viewpoint, finite-width geometries with disconnected boundaries, such as a strip, provide the simplest setting to study how locality and boundary conditions determine the boundary dynamics of a gauge theory in the presence of two edges. In such setups, the prevailing expectation is that counter-propagating chiral modes appear at opposite boundaries \cite{Tong:2016kpv,Susskind:2001}.
This expectation is supported both by semiclassical intuition (e.g.\ skipping orbits) and by effective field theory descriptions of annulus and multi-edge geometries \cite{Wen:2004ym}.
 However, in most analyses the boundary conditions (BC) are either fixed in a phenomenological way or the discussion effectively reduces to a single boundary. To the best of our knowledge, a fully explicit derivation of a pair of counter-propagating Tomonaga-Luttinger edge theories on a finite strip, starting from the most general local quadratic boundary action in the sense of Symanzik \cite{Symanzik:1981wd}, has not been provided. This leaves a gap between intuitive phenomenological arguments and a controlled, local field-theoretic formulation of CS theory with two boundaries.
A finite Hall strip is widely expected to support two counter-propagating chiral edge modes, with equal and opposite chiral velocities. In the condensed-matter literature this picture is typically established by combining an explicit edge-theory derivation on one boundary with additional, model-dependent or semiclassical input to infer the dynamics on the opposite edge, for instance through a particular background-field gauge choice for the vector potential of a uniform magnetic field together with a confining potential, or via skipping-orbit arguments. The purpose of the present work is to show that, in abelian CS theory on a strip, the two-edge structure can be obtained from a minimal and genuinely field-theoretic set of principles. We rely only on ingredients intrinsic to quantum field theory, namely locality in the Symanzik sense, the topological character of the bulk theory, and its gauge symmetries, and we use them to formulate the theory with two disconnected boundaries and to determine the associated boundary dynamics. In this way, the appearance of two counter-propagating chiral sectors and the constraint relating their propagation speeds are traced back to general QFT principles rather than to external microscopic assumptions.
Filling this gap is not merely a technical exercise: it is essential for understanding edge dynamics in finite geometries, where interactions, disorder, and inter-edge couplings, for instance through Casimir-type effects \cite{Casimir:1948dh,Blasi:1992mm,Milton:2001}, may play a fundamental role. In this work, we present a systematic construction that makes the appearance of counter-propagating edge modes on a strip fully explicit. By carefully analyzing the possible BC for the CS field and studying the resulting boundary dynamics, we show that modes with opposite chiralities emerge naturally at opposite edges. Our approach is based entirely on conventional quantum field theory techniques, following Symanzik's prescription for introducing boundaries in local quantum field theory, and is consistent with the bulk topological structure. To our knowledge, this represents the first explicit Symanzik-type derivation of such edge states in a controlled and fully field-theoretic manner.
Beyond its immediate application to quantum Hall systems, our framework lends itself to other physical scenarios described by CS-like theories \cite{Maggiore:2018bxr,Bertolini:2020hgr,Bertolini:2021iku,Maggiore:2019wie}. A particularly interesting example arises in hydrodynamic systems, notably the ``shallow water'' gauge theory discussed in \cite{Tong:2022gpg}. In that context one considers a layer of fluid of finite depth, bounded below by a rigid bottom at $x^2=0$ and above by a free surface at $x^2=h$ supporting long-wavelength surface waves. In our language, the strip $0\leq x^2\leq h$ can be interpreted precisely as such a shallow channel, with $x^1$ playing the role of the propagation direction of the waves and the two boundaries modeling the bottom and the free surface. Translating the physical constraints (no flow through the bottom and freely propagating deformations of the surface) into BC on the gauge field, our formalism captures the universal, non-dissipative edge dynamics of these setups. At the same time, only the topological and kinematic aspects are encoded: dissipative effects, detailed equations of state and other microscopic properties of concrete shallow water systems lie beyond the present analysis and would require additional, model-dependent input. This extension illustrates the versatility of our approach: by modifying the nature of the BC, the same underlying topological theory can describe both electronic and fluid systems, revealing deep structural parallels.
These examples motivate a systematic field-theoretic treatment of CS theory on finite geometries with multiple boundaries.
While edge physics in CS theories and quantum Hall systems has been extensively discussed in the literature,
including annulus or cylinder geometries and multi-edge setups \cite{Wen:1992vi,Stone:1990iw,Fradkin:1991wy,Zhang:1992eu,Wen:2004ym},
these studies typically either fix the BC in a phenomenological way or focus effectively on a single boundary.
Related field-theoretic analyses of CS theories with nontrivial boundaries, defects and interfaces can be found in \cite{Kapustin:2010,Fliss:2017wop,Fliss:2020,Gutperle:2019}.
However, the focus in these works is typically on topological boundary conditions, entanglement-related observables, or AdS$_3$/CFT$_2$ interface constructions, rather than on an explicit Symanzik-type derivation of two
counter-propagating Tomonaga-Luttinger edge modes on a finite strip, which is the goal of the present analysis.
In contrast, we provide a fully field-theoretic derivation of two counter-propagating edge modes in abelian CS gauge theory on a strip $0\leq x^2\leq h$, by combining Symanzik's method for implementing boundaries in local quantum field theory with the topological structure of the bulk theory. We construct the most general local quadratic boundary action consistent with locality and power counting, and classify the resulting boundary conditions that yield chiral Kac-Moody algebras on each edge. We then show how these algebras give rise to a pair of 
Tomonaga-Luttinger actions for edge modes with velocities $v$ and $\bar v$, related by a flip symmetry. An important outcome of the construction is that the edge velocities are independent of the strip width $h$: they are entirely controlled by the CS coupling and by boundary parameters, rather than by geometric details of the bulk region, as expected due to its topological nature.\\

This paper is organized as follows. Section 2 introduces the field-theoretic setup on a strip, formulates the bulk action and the most general local boundary action, and derives the resulting equations of motion and boundary conditions. We then obtain the broken gauge Ward identity in the presence of boundaries and show that it implies
chiral Kac-Moody current algebras with opposite central charges on the two edges, together with the corresponding boundary degrees of freedom described by scalar fields. Section 3 derives the induced 2D boundary actions and the bulk-boundary matching that relates the boundary parameters to the physical edge velocities, including the conditions under which a flip symmetry enforces $\bar v=-v$. Finally, Section 4 summarizes our results
and discusses possible extensions and applications of the framework.\\

{\bf Notations}

We work in $2+1=3$D spacetime dimensions. Greek indices $\mu,\nu,\rho,\ldots = \{0,1,2\}$ label
bulk coordinates, while Latin indices $i,j,k,\ldots = \{0,1\}$ refer to coordinates along the
edges. The Minkowski metric is $\eta_{\mu\nu} = \mathrm{diag}(-1,1,1)$. The strip extends in the
transverse direction $x^2$ with $0 \le x^2 \le h$. We denote boundary coordinates at the lower and
upper edges by
\begin{equation}
  X = (x^0,x^1,0)\,, \qquad \bar X = (x^0,x^1,h)\,.
\end{equation}
The Levi-Civita symbol is normalized as $\epsilon_{012} = 1 = -\epsilon^{012}$, and we define
$\epsilon^{ij2} = \epsilon^{2ij} \equiv \epsilon^{ij}$.

\section{Chern-Simons theory on a strip}\label{sec:bulk}

\subsection{Action, equations of motion, and boundary conditions}

We consider the CS theory defined on the strip $0\leq x^2\leq h$, where the presence of the boundaries at $x^2=0$ and $x^2=h$,  is implemented by two Heaviside step distributions $\theta(x^2)$ and  $\theta(h-x^2)$. The theory is therefore described by the total action
	\be
	S_{tot}=S_{\textsc{cs}}+S_{gf}+S_J+S_{bd}\ ,
	\ee
where
	\be\label{Scs}
	S_{\textsc{cs}}= \frac{\kappa}{2} \int d^3x \ \theta(x^2) \theta(h-x^2)\epsilon^{\mu \nu \rho}  A_\mu \partial_\nu A_\rho 
	\ee
is the CS action with ``coupling'' constant $\kappa$. Recall that $\kappa$ is a positive constant
\be
\kappa>0\ ,
\ee
since it is related to the filling factor $\nu$ of the Landau levels in the quantum Hall effect \cite{Tong:2016kpv} and to the winding number of Wilson loops \cite{Witten:1988hf}.
The gauge fixing term, corresponding to the axial gauge condition
\be
A_2=0\ ,
\label{A2=0}\ee
is given by
\be
	S_{gf}=\int d^3x \ \theta(x^2) \theta(h-x^2) b A_2\ ,
\label{Sgf}	\ee
where $b(x)$ acts as a Lagrange multiplier \cite{Nakanishi:1966zz,Lautrup:1967zz,Gambuti:2020onb}. 
External sources are introduced through
	\be
	S_J=\int d^3x \ \theta(x^2) \theta(h-x^2) J^i A_i \ ,
	\ee
where $J^i(x)$ couples to the gauge field $A_i(x)$. Finally, $S_{bd}$ denotes the most general local, quadratic boundary term consistent with locality and power counting in the sense of Symanzik \cite{Symanzik:1981wd}, localized at $x^2=0$ and $x^2=h$:	
	\bea
	S_{bd} &=& S^{(0)}_{bd} + S^{(h)}_{bd} \nonumber \\
		&=& \int d^3x\; \delta(x^2)\tfrac1 2 a^{ij}A_iA_j+\int d^3x\;\delta(x^2-h)\tfrac1 2 b^{ij}A_iA_j\ ,  \label{Sbd}
\eea
where $a^{ij}$ and $b^{ij}$ are symmetric, dimensionless constant matrices
	\be\label{ab}
	a^{ij}=a^{ji}\equiv\left(
		\begin{array}{cc}
		a_0&a_1\\
		a_1&a_2
		\end{array}\right)\quad;\quad
	b^{ij}=b^{ji}\equiv\left(
		\begin{array}{cc}
		b_0&b_1\\
		b_1&b_2
		\end{array}\right)\ .
	\ee
From a physical point of view, the matrices $a^{ij}$ and $b^{ij}$ parametrise the most general linear and local response of the two boundaries to the gauge field, in the sense of Symanzik's construction of boundary actions in quantum field theory \cite{Symanzik:1981wd}. The boundary term $S_{bd}$ is quadratic in the components $A_0$ and $A_1$ evaluated at $x^2=0$ and $x^2=h$, so that the entries $a_0,a_1,a_2$ (and similarly $b_0,b_1,b_2$) weight the combinations $A_0^2$, $A_0A_1$ and $A_1^2$ on each edge. In effective field theory language, these coefficients play the role of boundary couplings encoding short distance physics in a thin layer near the edge: they determine how costly it is, in energy, to excite temporal and longitudinal components of the gauge potential at the boundary \cite{Diehl:1997}.
In particular, the coefficients $a_2$ and $b_2$, which multiply $A_1^2$, control the response of the system to deformations of the gauge field along the direction of the edge and will later be seen to determine the coefficients of the Tomonaga-Luttinger actions describing the edge modes. Large values of $|a_2|$ or $|b_2|$ correspond to ``hard'' or rigid boundaries, where longitudinal distortions of the edge fields are strongly suppressed, while smaller values describe ``softer'' boundaries that can be more easily deformed. The off-diagonal entries $a_1$ and $b_1$ instead mix $A_0(x)$ and $A_1(x)$, i.e.\ they couple the temporal and longitudinal
boundary components of the gauge potential and encode a cross-response between the boundary charge-density and the current along the edge \cite{Wen:1992vi,Stone:1990iw}.
Similar boundary couplings appear in other analyses of topological and gauge theories with boundaries. In the Symanzik-inspired approach to topological field theories \cite{Amoretti:2014iza,Maggiore:2018bxr}, one also writes the most general local boundary action with free coefficients that encode the short distance physics near the edge. In abelian CS theory and at interfaces between different topological phases \cite{Kapustin:2010,Fliss:2017wop,Fliss:2020}, the choice of such boundary couplings controls which combinations of edge modes remain relevant at low energies and which are pushed to higher energies and do not contribute to the long distance dynamics.
In a microscopic quantum Hall realisation, the effective parameters $(a_{ij},b_{ij})$ would be generated by integrating out the bulk degrees of freedom in the presence of a given confining potential and electron-electron interactions, so that different regions of the $(a_{ij},b_{ij})$ space correspond to different classes of physical edges (hard-wall vs.\ soft-wall profiles, screened vs.\ unscreened boundaries, etc.) \cite{Chamon:1993,Chklovskii:1992}. 
The boundary term $S_{bd}$ thus follows the general Symanzik prescription for implementing boundaries in quantum field theory \cite{Symanzik:1981wd} and has been used in several topological and gauge theoretic contexts \cite{Amoretti:2014iza,Maggiore:2018bxr,Kapustin:2010}. The most general BC are then derived from $S_{bd}$ by varying the total action to obtain the equations of motion
\bea
	\frac{\delta S_{tot}}{\delta A_i}&=& \theta(x^2) \theta(h-x^2) \left[
	\kappa \epsilon^{ij}(\partial_jA_2-\partial_2A_j) + J^i  \right] +
	\nonumber\\
	&& \quad + \left[\delta(x^2) \left( a^{ij}- \tfrac{\kappa}{2} \epsilon^{ij}\right)+\delta(x^2-h) \left( b^{ij}+\tfrac{\kappa}{2} \epsilon^{i j} \right)\right]  A_j\label{EoMa}\\
	\frac{\delta S_{tot}}{\delta A_2}&=& \theta(x^2) \theta(h-x^2) \left( \kappa \epsilon^{ij} \partial_i A_j + b \right) \ .
\eea
To extract the boundary conditions we integrate Eq.~\eqref{EoMa} over small intervals $[0,\epsilon]$ and $[h-\epsilon,h]$ in $x^2$ and then take $\epsilon \to 0$. The bulk terms, proportional to $\theta(x^2)\theta(h-x^2)$, vanish in this limit, and only the $\delta$-supported boundary contributions survive.
Evaluating these equations on-shell near the  boundaries gives
	\begin{align}
	\lim_{\epsilon\to0}\int^\epsilon_0dx^2\frac{\delta S_{tot}}{\delta A_i}=0\quad\Rightarrow\quad& \left( a^{ij}- \tfrac{\kappa}{2} \epsilon^{ij}\right)A_j(X)=0\label{BC0}\\
	\lim_{\epsilon\to0}\int^h_{h-\epsilon}dx^2\frac{\delta S_{tot}}{\delta A_i}=0\quad\Rightarrow\quad&\left( b^{ij}+\tfrac{\kappa}{2} \epsilon^{i j} \right)A_j(\bar X)=0\label{BCh}\ ,
	\end{align}
whose components explicitly read
	\begin{align}
	i=0\ :\qquad a_0A_0(X)+\left(a_1+\tfrac \kappa 2 \right)A_1(X)=0&\label{bc1}\\
			b_0A_0(\bar X)+\left(b_1-\tfrac \kappa 2 \right)A_1(\bar X)=0&\label{bc2}\\
	i=1\ :\qquad\left(a_1- \tfrac \kappa 2 \right)A_0(X)+a_2A_1(X)=0&\label{bc3}\\
		\left(b_1+\tfrac \kappa 2 \right)A_0(\bar X)+b_2A_1(\bar X)=0&\ .\label{bc4}
	\end{align}
	The relative minus sign between the two boundary contributions at $x^2=0$ and $x^2=h$ ultimately reflects the
opposite orientation of the two edges (equivalently, the opposite outward normals at $x^2=0$ and $x^2=h$).
This sign is responsible for the opposite central terms in the boundary current algebras, and hence for
the opposite chiralities of the two edge modes.
These BC admit several classes of solutions involving the boundary fields $\{A_i(X);A_i(\bar X)\}$ and the constant matrices $\{a^{ij};b^{ij}\}$. Among these, we focus on those that allow a consistent holographic mapping between the 3D bulk theory and the induced 2D theory, as analyzed in Section 3.

\subsection{Broken Ward identity and boundary degrees of freedom}

From the on-shell EoM \eqref{EoMa} and the BC \eqref{BC0} and \eqref{BCh}, we get the following Ward identity, broken by the boundaries
	\be
	\int^h_0dx^2\partial_iJ^i=
	\left.\kappa\epsilon^{ij}\partial_i A_j\right|_{x^2=h}
	-\left.\kappa\epsilon^{ij}\partial_i A_j\right|_{x^2=0}\label{WIsep}\ .
	\ee
Setting $J^i(x)=0$ ($i.e.$ going on-shell) gives the constraint
	\be
	\left.\epsilon^{ij}\partial_i A_j\right|_{x^2=h}
	-
	\left.\epsilon^{ij}\partial_i A_j\right|_{x^2=0}=0\label{cc}\ .
	\ee
Since this relation involves field values at separate points, it is explicitly nonlocal along the transverse direction. In principle, one could envisage more general BC in which the gauge fields at $x^2=0$ and $x^2=h$ are directly correlated, so that the constraint \eqref{cc} is realized in a genuinely nonlocal way. 
For instance, one could add explicitly inter-edge couplings through bilocal terms in the boundary action,
\begin{equation}
S_{\rm bd}^{\rm nonloc}\;\sim\;\int d^2X\, d^2\bar X\; \mathcal{K}(X-\bar X)\, A_i(X)\,A_j(\bar X)\,,
\label{eq:Sbd_nonloc}
\end{equation}
or enforce boundary constraints that relate the fields at $x^2=0$ and $x^2=h$ directly. In such a setup,
the Ward identity would constrain only the \emph{combined} expression in \eqref{cc}, and one should not expect two
strictly independent edge theories. In particular, effective edge observables could acquire a dependence on the
strip width $h$ once inter-edge correlations are allowed.
In the present work, however, we focus on $local$ BC: the boundary action is taken to be a sum of contributions localized at each edge, and the sources coupled to boundary operators on the two boundaries are treated as independent. Under these assumptions, the functional variations with respect to the boundary sources at $x^2=0$ and at $x^2=h$ are independent, and the Ward identity must hold for arbitrary such variations on each edge separately. This implies that the two boundary terms appearing in \eqref{cc} must vanish independently, and we are led to impose the local conditions
\be
	\left.\epsilon^{ij}\partial_i A_j\right|_{x^2=0}=0\quad\mbox{and}\quad\left.\epsilon^{ij}\partial_i A_j\right|_{x^2=h}=0\label{cc-indep}\ ,
	\ee
which can be solved by introducing scalar fields $\phi(X)$ and $\psi(\bar X)$ on the two boundaries
\be
	A_i(X)=\partial_i\phi(X)\quad\mbox{and}\quad A_i(\bar X)=\partial_i\psi(\bar X)\label{A=phi}\ ,
\ee
where $\phi(X)$ and $\psi(\bar X)$ are scalar boundary degrees of freedom living, respectively, on $x^2=0$ and $x^2=h$. Equations \eqref{cc-indep} should thus be viewed as a consistent local realization of the broken Ward identity on the strip, appropriate for two independent edge theories; more exotic, explicitly nonlocal couplings between the edges, although in principle possible, lie beyond the scope of the present analysis.

\subsection{Boundary Kac-Moody algebras}

Taking the functional derivative of the broken Ward identity \eqref{WIsep} with respect to $J^k(x')$ and setting the sources to zero, we obtain 
\be
\left.\frac{\delta}{\delta J^k(x')}\int dx^2 \partial_iJ^i \right|_{J=0}=
\kappa \epsilon^{ij}\partial_i
\frac{\delta}{\delta J^k(x')}
\left.\left(
\left. A_j  \right|_{x^2=h} -
\left. A_j  \right|_{x^2=0}
\right)\right|_{J=0} \ ,
\ee
which implies
\be
\int dx^2 \partial_i\delta^i_k\delta^{(3)}(x-x') =
\kappa\epsilon^{ij}\partial_i 
\left.\left(
\left.\frac{\delta^2Z_c}{\delta J^k(x')\delta J^j(x)}\right|_{x^2=h}
-
\left.\frac{\delta^2Z_c}{\delta J^k(x')\delta J^j(x)}\right|_{x^2=0}o
\right)\right|_{J=0}\ ,
\ee
and thus
\bea
\partial_k\left[\delta(x^0-x'^0)\delta(x^1-x'^1)\right] &=&
\kappa\epsilon^{ij}\partial_i\left\{
i\left.\left\langle T\left(A_k(x')A_j(x)\right)\right\rangle\right|_{x^2=h}
-
i\left.\left\langle T\left(A_k(x')A_j(x)\right)\right\rangle\right|_{x^2=0}
\right\}\nonumber\\
&=&
i\kappa\delta(x^0-x'^0)\left(
-\left.\left\langle\left[ A_1(x),A_k(x')\right]\right\rangle\right|_{x^2=h}
+
\left.\left\langle\left[ A_1(x),A_k(x')\right]\right\rangle\right|_{x^2=0}
\right)\nonumber\\
&=&
i\kappa\delta(x^0-x'^0)\left(
-\left\langle\left[ A_1(\bar X),A_k(x')\right]\right\rangle
+
\left\langle\left[ A_1(X),A_k(x')\right]\right\rangle
\right)\ .\label{pre-algebra}
\eea
Here, the generating functional $Z_c[J]$ of connected Green's functions satisfies \cite{Srednicki:2007qs}
	\bea
	\left.\frac{\delta Z_c}{\delta J^j(x)}\right|_{J=0}&\equiv&\langle A_j(x)\rangle\\
	\left.\frac{\delta^2 Z_c}{\delta J^j(x)\delta J^k(x')}\right|_{J=0}&\equiv &
	i\langle T[A_j(x)A_k(x')]\rangle\ ,
	\eea
and the time-ordered product $T$ is defined as
	\be
	\langle T[A_i(x)A_j(x')]\rangle\equiv\langle A_i(x)A_j(x')\theta(x^0-x'^0)+A_j(x')A_i(x)\theta(x'^0-x^0)\rangle\ .
	\ee
By setting in $k=1$ \eqref{pre-algebra}, we find the equal-time relation
\be
\left\langle\left[ A_1(\bar X),A_1(x')\right]\right\rangle
-
\left\langle\left[ A_1(X),A_1(x')\right]\right\rangle =
\frac{i}{\kappa}\partial_1\delta(x^1-x'^1)\ .
\label{algebratot}\ee
Observing that at $x^2=0$, only $A_i(X)$ exists while $A_i(\bar X)$ vanishes (and $viceversa$), at the two boundaries of the theory \eqref{algebratot} gives 
\bea
x^2=0 &:&
\left\langle\left[ A_1(X),A_1(X')\right]\right\rangle = -\frac{i}{\kappa}\partial_1\delta(x^1-x'^1) \label{KMx2=0}\\
x^2=h &:&
\left\langle\left[ A_1(\bar X),A_1(\bar X')\right]\right\rangle = \frac{i}{\kappa}\partial_1\delta(x^1-x'^1) \ .\label{KMx2=h}
\eea
These represent two independent Kac-Moody algebras \cite{Kac:1967jr,Moody:1966gf} located at the boundaries
$x^2=0$ and $x^2=h$ at equal times $x^0=x'^0$, with opposite central charges  $\mp\frac{1}{\kappa}$. We remark that, while we perform the derivation in the axial gauge \eqref{A2=0} for technical convenience, the appearance of the boundary Kac-Moody central terms is tied to the broken gauge Ward identity and is therefore a universal feature of the boundary algebra.

\section{Bulk-boundary correspondence}

The two boundary algebras \eqref{KMx2=0} and \eqref{KMx2=h}, when expressed in terms of the 2D scalar degrees of freedom $\phi(X)$ and $\psi(\bar X)$ \eqref{A=phi} embed the canonical commutation relations
	\begin{align}
	\left[\partial_1\phi\ ,\ \partial'_1\phi'\right]&=-\frac{i}{\kappa} \partial_1\delta(x_1-x'_1)&&\longrightarrow&&\left[\,q\ ,\ p'\,\right]=i\delta(x_1-x_1')\qquad\\
	\left[\partial_1\psi\ ,\ \partial'_1\psi'\right]&=\frac{i}{\kappa} \partial_1\delta(x_1-x'_1)&&\longrightarrow&&\left[\,\bar q\ ,\ \bar p' \,\right]=i\delta(x_1-x_1')\ , 
	\end{align}
where $\partial'_1\equiv \frac{\partial}{\partial x'^{1}}$, $q=q(X),\ p=p(X),\ \bar q=\bar q(\bar X)$ and $\bar p=\bar p(\bar X)$, once we identify
\begin{align}
	q\equiv\phi\quad&;\quad p\equiv -\kappa\partial_1\phi\label{pq}\\
	\bar q\equiv\psi\quad&;\quad \bar p\equiv \kappa\partial_1\psi\ .\label{bpq}
	\end{align}
Since the definitions of the boundary degrees of freedom depend only on the derivatives of $\phi(x)$ and $\psi(x)$, they are in
particular invariant under 
the shift invariance
	\be
	\delta_{s}\phi=\eta\quad;\quad\bar\delta_{s}\psi=\bar\eta\ ,\label{shift}
	\ee
with $\eta$ and $\bar \eta$ constants. 
Since the Lagrangian and the Hamiltonian are related through the Legendre transformation
\be
L(q_i,\dot q_i,t)= p_i\dot q_i - H(q_i,p_i,t)\ ,
\ee
and recalling that the boundary fields $\phi(X)$ and $\psi(\bar X)$ are dimensionless, the most general 2D induced actions consistent with power counting and invariant under the shift symmetries \eqref{shift} take the form
\begin{align}
	S_{2D}&=\int d^2X\left[
	-\kappa\partial_1\phi\partial_0\phi-\alpha^2\left(\partial_1\phi\right)^2\right]\label{S2d}\\
	\bar S_{2D}&=\int d^2\bar X\left[\kappa\partial_1\psi\partial_0\psi-\beta^2\left(\partial_1\psi\right)^2\right]\ ,\label{barS2d}
	\end{align}
associated to the 2D Tomonaga-Luttinger Hamiltonians \cite{Haldane:1981zza,Wen:2004ym}
\bea
\left.H(\phi)\right|_{x^2=0} &=& \alpha^2\left(\partial_1\phi\right)^2 \label{Hx2=0}\\
\left.H(\psi)\right|_{x^2=h} &=& \beta^2\left(\partial_1\psi\right)^2\ ,\label{Hx2=h}
\eea
where $\alpha^2$ and $\beta^2$ are positive, dimensionless coefficients. Positivity of $\alpha^2$ and $\beta^2$ follows from stability: the Hamiltonians \eqref{Hx2=0} and \eqref{Hx2=h} are
bounded from below only for $\alpha^2>0$ and $\beta^2>0$, which also fixes the allowed sign branches for the
boundary couplings entering the holographic matching.
The on-shell EoM obtained from these actions are
	\begin{align}
x^2=0\ :\qquad 	\frac{\delta S_{2D}}{\delta\phi(X)}&=2\partial_1\left(\kappa\partial_0\phi(X)+\alpha^2\partial_1\phi(X)\right)=0\label{EoMphi}\\
x^2=h\ :\qquad 	\frac{\delta\bar S_{2D}}{\delta\psi(\bar X)}&=2\partial_1\left(-\kappa\partial_0\psi(\bar X)+\beta^2\partial_1\psi(\bar X)\right)=0\ .\label{EoMpsi}
	\end{align}
We thus arrive at one of the main results of this paper: the existence, on the two boundaries $x^2=0$ and $x^2=h$, of the Kac-Moody algebras \eqref{KMx2=0} and \eqref{KMx2=h} with equal and opposite central charges $\pm 1/\kappa$, induces the 2D actions $S_{2D}$ \eqref{S2d} and $\bar S_{2D}$ \eqref{barS2d}, which are of the Tomonaga-Luttinger type. Therefore, they describe, through their EoM \eqref{EoMphi} and \eqref{EoMpsi}, chiral bosonic edge modes propagating along the boundaries with $opposite$ velocities
\be
v\equiv\frac{\alpha^2}{\kappa}\ ,
\ee
and
\be
\bar v\equiv-\frac{\beta^2}{\kappa}\ ,
\ee
which depend on the CS ``coupling'' constant $\kappa$ and the positive parameters $\alpha^2$ and $\beta^2$. We will now relate these coefficients to the parameters appearing in $S_{bd}$ \eqref{Sbd} through a holographic matching condition. To do so, we rewrite
the BC \eqref{bc1}--\eqref{bc4} in terms of the scalar fields \eqref{A=phi}
\begin{align}
	x^2=0\ :\qquad 
a_0\partial_0\phi(X)+\left(a_1+\tfrac \kappa 2 \right)\partial_1\phi(X)=0&\label{bc1phi}\\
\qquad\left(a_1- \tfrac \kappa 2 \right)\partial_0\phi(X)+a_2\partial_1\phi(X)=0&\label{bc3phi}\\
	x^2=h\ :\qquad	
b_0\partial_0\psi(\bar X)+\left(b_1-\tfrac \kappa 2 \right)\partial_1\psi(\bar X)=0&\label{bc2psi}\\
\left(b_1+\tfrac \kappa 2 \right)\partial_0\psi(\bar X)+b_2\partial_1\psi(\bar X)=0&\ .\label{bc4psi}
	\end{align}
Notice that 
\be
\partial_i\phi(X)=\partial_i\psi(\bar X)=0\ ,
\ee
which correspond to the Dirichlet BC
\be
A_i(X)=A_i(\bar X)=0\ ,
\ee
trivially satisfy \eqref{bc1phi}--\eqref{bc4psi}, thereby annihilating the boundary actions $S_{2D}$ \eqref{S2d} and $\bar S_{2D}$  \eqref{barS2d}.
The bulk-boundary holographic contact is realized by requiring that the EoM \eqref{EoMphi} and \eqref{EoMpsi} be compatible with the BC \eqref{bc1phi}--\eqref{bc4psi}. This compatibility can be achieved by substituting  \eqref{bc1phi} and \eqref{bc3phi} into \eqref{EoMphi} for the boundary $x^2=0$, and \eqref{bc2psi} and \eqref{bc4psi} into \eqref{EoMpsi} for $x^2=h$. 

\begin{itemize}

\item \emph{Boundary at $x^2=0$.}
The nontrivial solutions are
\be
\alpha^2= - a_2>0 \quad ; \quad
a_0 =0 \quad ; \quad
a_1 = - \frac{\kappa}{2} \quad ; \quad
a_2< 0\ ,
\label{solx2=0.1}\ee
or
\be
\alpha^2=\kappa\frac{a_1+\kappa/2}{a_0}>0 \quad ; \quad
a_0\neq 0 \quad ; \quad
a_1\neq-\frac{\kappa}{2} \quad ; \quad
a_2=\frac{a_1^2-\kappa^2/4}{a_0}\ .
\label{solx2=0.2}\ee
Hence, the 2D action $S_{2D}$ \eqref{S2d} that realizes this bulk-boundary correspondence is
\be
S_{2D}=\int d^2X\left[
-\kappa\partial_1\phi\partial_0\phi +a_2 \left(\partial_1\phi\right)^2\right]\ ,\label{S2dfinal}
\ee
whose on-shell EoM 
\be
	\partial_1\left(\partial_0\phi+v\partial_1\phi\right)=0
\label{eomv}\ee
describes a chiral boson moving on $x^2=0$ with velocity
\be
v\equiv -\frac{a_2}{\kappa}>0\ ,
\label{v}\ee
which may be fixed phenomenologically, providing a physical interpretation for the parameter $a_2$ appearing in $S^{(0)}_{bd}$ \eqref{Sbd}. In fact equation \eqref{eomv} represents a continuity equation for the density  $\rho(X)$ and the current $J(X)$, defined as
	\be
	\rho \equiv \partial_1\phi\quad;\quad     J\equiv v\rho\ ,
\label{conteqx2=0}	\ee
so that 
\be
	\partial_0\rho+\partial_1J=0\ .
\ee
The alternative solution \eqref{solx2=0.2} is physically equivalent to \eqref{solx2=0.1}, describing a chiral boson with positive velocity. Requiring the Hamiltonian \eqref{Hx2=0} to be bounded from below enforces $\alpha^2>0$ and hence $a_2<0$ in this branch.

\item \emph{Boundary at $x^2=h$.} The situation at the other boundary $x^2=h$  is completely analogous. Compatibility is achieved for
\be
\beta^2= - b_2>0 \quad ; \quad
b_0 =0 \quad ; \quad
b_1 =  \frac{\kappa}{2} \quad ; \quad
b_2< 0\ ,
\label{solx2=h.1}\ee
or
\be
\beta^2=-\kappa\frac{b_1-\kappa/2}{b_0}>0 \quad ; \quad
b_0\neq 0 \quad ; \quad
b_1\neq\frac{\kappa}{2} \quad ; \quad
b_2=\frac{b_1^2-\kappa^2/4}{b_0}\ .
\label{solx2=h.2}\ee
From \eqref{solx2=h.1}, the 2D action $\bar S_{2D}$ \eqref{barS2d} consistent with holographic contact is
\be
\bar S_{2D}=\int d^2\bar X\left[
\kappa\partial_1\psi\partial_0\psi +b_2
\left(\partial_1\psi\right)^2\right]\ ,\label{barS2dfinal}
\ee
which describes a chiral boson propagating at $x^2=h$ with negative velocity
\be
\bar v\equiv \frac{b_2}{\kappa}<0\ ,
\label{barv}\ee
thereby determining the parameter $b_2$  in $S_{bd}$ \eqref{Sbd}. As for the previous case, both \eqref{solx2=h.1} and \eqref{solx2=h.2} describe the same physical scenario: a chiral boson on $x^2=h$ moving with negative velocity. The stability condition $\beta^2>0$ implies $b_2<0$ in the solution \eqref{solx2=h.1}.
\end{itemize}
From the point of view of the effective 2D description, the coefficients $\alpha^2$ and $\beta^2$ in the boundary actions \eqref{S2d} and \eqref{barS2d} play the role of (edge) stiffness parameters for the chiral bosons: they multiply the spatial-gradient terms $(\partial_1\phi)^2$ and $(\partial_1\psi)^2$ and therefore control the energy cost of modulating the edge fields along $x^1$.
Together with the mixed chiral term proportional to $\kappa\,\partial_1\phi\,\partial_0\phi$ (and $\kappa\,\partial_1\psi\,\partial_0\psi$), they also fix the propagation velocities of the edge modes (for fixed $\kappa$, larger $\alpha^2$ or $\beta^2$ implies larger $|v|$ and $|\bar v|$), as in the standard chiral-boson/chiral-Luttinger-liquid effective description of quantum Hall edges \cite{Tong:2016kpv,Wen:1990se,Stone:1990iw}.
Finally, imposing compatibility between the bulk equations of motion, the boundary conditions derived from $S_{bd}$, and the edge equations of motion \eqref{EoMphi}, \eqref{EoMpsi} expresses $\alpha^2$ and $\beta^2$ in terms of the boundary parameters $a_{ij}$ and $b_{ij}$.
In the simplest branch of solutions, Eqs.~\eqref{solx2=0.1} and \eqref{solx2=h.1} imply in particular
\begin{equation}
\alpha^2=-a_2>0\,, \qquad \beta^2=-b_2>0\,,
\end{equation}
so that the edge actions describe chiral modes with velocities \eqref{v} and \eqref{barv} 
\begin{equation}
 v=-\frac{a_2}{\kappa}>0 \quad \text{at } x^2=0\,, \qquad
 \bar v=\frac{b_2}{\kappa}<0 \quad \text{at } x^2=h\,,
\end{equation}
Thus the velocities $v$ and $\bar v$ are not free phenomenological inputs: they compactly encode the underlying boundary couplings collected in $S_{bd}$ and therefore reflect the microscopic properties of the edge, such as the effective rigidity of the confining profile and the strength of interactions in the edge region. In a microscopic quantum Hall realisation, these effective parameters are expected to be renormalised by the detailed shape of the confining potential and by electron-electron interactions, and can in practice be tuned by changing gate voltages, screening or edge reconstruction \cite{Chamon:1993}. In our language, such changes correspond to moving in the parameter space of the boundary matrices $(a_{ij},b_{ij})$, and hence of $(\alpha^2,\beta^2)$.\\

In summary, the two boundaries $x^2=0$ and $x^2=h$ support edge modes propagating with opposite velocities $v$ and $\bar v$ 
\be
v\bar v<0\ ,
\ee
unless Dirichlet BC are imposed on one, or both, of the two boundaries. This last case is particularly relevant in the ``shallow waters'' case, where only one mode propagates, as described in \cite{Tong:2022gpg}.\\

The CS term violates both parity and time reversal symmetry. The introduction of boundaries, implemented by the two $\theta$-functions in $S_{\textsc{cs}}$ \eqref{Scs}, also breaks translations and rotation invariance. However, the theory remains invariant under the flip symmetry exchanging the two boundaries $x^2=0$ and $x^2=h$, realized by combining a rotation by $\pi$ with a translation of $h$ along $x^2$ 
\be
x^0\rightarrow x^0\ \ ;\ \ 
x^1\rightarrow -x^1\ \ ;\ \ 
x^2\rightarrow -x^2+h\ ,
\label{flipx}\ee
which induces on the gauge fields the transformation
\be
A_0\rightarrow A_0 \ \ ;\ \
A_1\rightarrow -A_1 \ \ ;\ \
A_2\rightarrow -A_2\ .
\label{flipA}\ee
Imposing invariance under this flip symmetry requires setting
\be
a_0=b_0 \ \ ;\ \ 
a_1=-b_1 \ \ ;\ \ 
a_2=b_2 \ ,
\label{flipcoeff}\ee
with the consequence that the edge modes at $x^2=0$ and $x^2=h$ move with {\it equal and opposite} velocities
\be
\bar v = - v\ .
\ee
The standard semiclassical way to understand that the chiral edge mode velocities in quantum Hall systems are equal in magnitude and opposite in sign is to use the picture of ``skipping orbits'' \cite{Halperin:1982,Buttiker:1988}: electrons that in the bulk move in cyclotron orbits under a perpendicular magnetic field can no longer complete full orbits at the edges and are therefore forced to bounce along the boundary, producing an effective edge current.
This picture is phenomenologically described as follows.
In the Landau gauge, with the vector potential chosen so as to preserve translational invariance along the strip, the momentum $k_1$ conjugate to $x^1$ is a good quantum number, and the guiding-center coordinate in the transverse direction $x^2$ is
\begin{equation}
\bar x^2= k_1 \,\ell_B^2\ ,
\end{equation}
with $\ell_B$ the magnetic length.
A confining potential $V(\bar x^2)$ lifts the Landau levels $E_n$ near the boundaries, producing an effective dispersion relation
\begin{equation}
E(k_1) = E_n + V(\bar x^2)\ .
\end{equation}
The corresponding group velocity of the edge state along $x^1$ is then
\begin{equation}
v_1 = \frac{\partial E}{\partial k_1}
      = \ell_B^2\;\partial_2 V\ .
\end{equation}
This Landau-gauge derivation of the edge dispersion and velocity is standard in the quantum Hall literature \cite{Wen:1992vi,Tong:2016kpv}. If the two edges of the sample are modeled by a symmetric potential, such that the confinement slopes $\partial_2 V$ have equal magnitude and opposite sign at the two boundaries, the resulting edge modes propagate with velocities that satisfy $\bar v = -v$: equal in magnitude, opposite in direction.
This reasoning, however, depends explicitly on the assumed external potential, which is an input introduced by hand in the standard treatment of the quantum Hall effect.
In contrast, in the present framework the same conclusion arises \emph{without} any external confining potential:
it follows purely from gauge symmetry, more precisely from the breaking of gauge invariance induced by the presence of the two spatial boundaries.
The equality and opposite sign of the edge velocities thus emerge as a gauge theoretic consequence of the boundary structure of the theory, rather than from model dependent assumptions about the microscopic confinement.
This highlights one of the conceptual strengths of the present approach, in which the standard phenomenological input (the confining potential) is replaced by a result stemming from the gauge symmetry itself.

\section{Summary and outlook} \label{sec-conclusion}

In this work we studied abelian Chern-Simons theory on a strip with two spatial boundaries and showed how the broken gauge Ward identity encodes the emergence of dynamical edge degrees of freedom. For suitable boundary conditions,
the boundary fields realize two independent chiral Kac-Moody algebras with opposite central charges, and the induced dynamics is captured by a pair of Tomonaga-Luttinger actions describing counter-propagating modes. In a symmetric setup, a discrete flip symmetry exchanging the two boundaries enforces $\bar v=-v$, so that the equality and opposite
sign of the edge velocities follow directly from gauge symmetry and boundary structure rather than from model-dependent assumptions about confining potentials. Therefore the full two-boundary system exhibits a global parity-like symmetry that compensates for the intrinsic parity violation of the bulk Chern-Simons term.  
The gauge-invariant description of the boundary modes in terms of the scalar fields $\phi(X)$ and $\psi(\bar X)$ leads to two chiral bosonic theories localized at $x^2=0$ and $x^2=h$, whose induced 2D dynamics is of Tomonaga-Luttinger type. The matching between the bulk boundary conditions and the boundary equations of motion defines a holographic-like bulk-boundary correspondence that fixes the allowed parameter combinations and relates them to physically measurable quantities, such as the edge velocities.
\\

It is instructive to compare our derivation with the standard argument often used in the quantum Hall literature to motivate counter-propagating edge modes with velocities of equal magnitude and opposite sign. In the conventional picture one chooses a convenient background-field gauge for the vector potential of a uniform magnetic field and introduces a smooth confining electrostatic potential so that the Landau levels bend near the boundaries. By linearizing the resulting edge dispersions, one
infers that low-energy excitations propagate along the two edges with opposite group velocities
\cite{Wen:1992vi,Stone:1990iw,Fradkin:1991wy,Halperin:1982}. While this construction is physically transparent, it relies on semiclassical and model-dependent inputs that are external to the topological Chern-Simons effective field theory. In our framework, by contrast, the qualitative structure of counter-propagating modes follows directly from general quantum field-theoretic principles: gauge invariance, locality and power counting fix the most general quadratic
boundary action, and consistency with the bulk equations of motion and the broken Ward identity selects the admissible boundary conditions. The edge theories and their chiral velocities are therefore determined without imposing any particular potential. At the same time, the set of allowed boundary conditions furnished by the theory is flexible enough to accommodate not only the symmetric case with two counter-propagating modes of equal speed (realized when the flip symmetry is enforced), but also configurations with opposite yet unequal velocities, as well as situations where Dirichlet conditions freeze one boundary while a single chiral mode propagates along the other.\\

Another outcome of our construction is that, within the local and decoupled-edge framework adopted here, the physical results do not depend on the strip width $h$. As long as the edges can be treated as effectively one-dimensional, this is consistent with the phenomenology of quantum Hall devices, where varying the sample width does not affect the local edge dynamics unless one resolves the finite transverse structure of the edge region itself \cite{Chamon:1993}. From the
field-theoretic point of view, the absence of any $h$-dependence is nontrivial: once a dimensionful parameter such as $h$ is introduced into a scale-invariant theory like Chern-Simons, one might generically expect observables to acquire a dependence on it. In our case this does not happen: the chiral velocities are controlled by the central charges of the boundary Kac-Moody algebras, fixed by the Chern-Simons coupling, while the geometric parameter $h$ drops out. This
$h$-independence should be understood as a property of the low-energy topological description: in more microscopic models, where additional dynamical scales are present or the internal structure of the edge region is resolved, subleading corrections with a residual dependence on $h$ may in principle appear, but such effects lie beyond the
scope of the present analysis.\\

While the existence of chiral edge modes in abelian Chern-Simons theory is well known, the strip geometry with two disconnected boundaries raises a genuine two-boundary issue that is usually left implicit: without further assumptions, the two edges may in principle be correlated and the emergence of two \emph{independent} edge sectors is not automatic. The main conceptual novelty of the present work was to make this point precise and to resolve it within a minimal and fully local QFT setting. Concretely, we worked in Symanzik's framework and imposed locality and power counting on the admissible boundary terms; together with gauge symmetry, implemented through the broken Ward identity, this fixed the most general quadratic boundary action and selected consistent boundary conditions. As a consequence, the theory on a strip supports two independent boundary Kac-Moody algebras, one per edge, and hence two counter-propagating chiral Tomonaga-Luttinger sectors. In this sense, the second edge mode is not inferred from a single-boundary analysis supplemented by external input, but follows from the two-boundary formulation under locality. This provides a controlled field-theoretic foundation for the standard quantum Hall expectation, which is often motivated by additional microscopic or semiclassical ingredients that are external to the topological Chern-Simons effective description. Relaxing locality by allowing genuine inter-edge couplings, for instance bilocal boundary terms, would generically correlate the two boundaries and may obstruct the appearance of two strictly independent algebras and modes; exploring such nonlocal extensions is a natural direction for future work.
\\

A related but conceptually distinct source of inter-boundary correlations arises when the bulk manifold has a noncontractible cycle, as in an annulus (or cylinder) geometry. In that case, the topological bulk theory admits Wilson loops with nontrivial holonomy, and gauge invariance requires that this global holonomy be consistently realized in the boundary description. This induces a natural nonlocal constraint that identifies the corresponding zero-mode sectors of the two boundary currents, even in the absence of explicit inter-edge interaction terms in the boundary action. This effect should be distinguished from the explicitly nonlocal inter-edge couplings mentioned above ($e.g.$\ bilocal boundary terms): it reflects global topological data and primarily constrains the zero modes, whereas the independence of the local propagating sectors in the strip setup follows within the local and decoupled-edge framework adopted here.\\

In conclusion, we have provided an explicit and self-consistent realization of bulk-boundary correspondence in abelian Chern-Simons theory on a strip with two boundaries. The 3D topological theory induces two 2D chiral theories with opposite central charges and opposite propagation velocities, offering a controlled setting to study the interplay among topological gauge dynamics, boundary conditions, and emergent edge degrees of freedom.  
Our results provide a starting point for more general constructions involving multiple interfaces, non-abelian extensions, and setups with explicit inter-edge couplings or coupled topological phases and boundary excitations~\cite{Amoretti:2013xya,Bertolini:2022sao,Bertolini:2023sqa,Bertolini:2023wie}.  \\


\section*{Acknowledgments}

We thank Dario Ferraro and Niccol\`o Traverso Ziani for clarifying discussions on the phenomenological field-theoretic description of Hall systems, and for their extremely helpful suggestions. M.D., C.M., and C.P. thank the Dublin Institute for Advanced Studies for hospitality and support during part of this work.

\end{document}